\documentclass[lettersize,hidelinks,conference]{IEEEtran}
\usepackage[english]{babel}
\usepackage{amsmath,amsfonts,amssymb, amsthm}
\usepackage{algorithmic}
\usepackage{cite} 

\newcommand{\algorithmicinitialize}{\textbf{Initialize:}}
\newcommand{\INITIALIZE}{\item[\algorithmicinitialize]}
\usepackage{algorithm}
\usepackage{array}
\usepackage[caption=true,font=footnotesize]{subfig}
\usepackage{textcomp}
\usepackage{stfloats}
\usepackage{url}
\usepackage{verbatim}
\usepackage{graphicx}
\def\BibTeX{{\rm B\kern-.05em{\sc i\kern-.025em b}\kern-.08em
T\kern-.1667em\lower.7ex\hbox{E}\kern-.125emX}}
\usepackage{balance}
\usepackage{orcidlink}
\usepackage{float}
\usepackage{lipsum}
\usepackage[acronym,shortcuts]{glossaries}
\theoremstyle{plain}

\graphicspath{{./Fig/}}
\usepackage{placeins}
\usepackage{textcomp}
\usepackage{multirow}

\newacronym{QoS}{QoS}{quality of service}
\newacronym{TX}{TX}{transmit}
\newacronym{RX}{RX}{receive}
\newacronym{IoT}{IoT}{Internet of Things}
\newacronym{SDR}{SDR}{semi-definite relaxation}
\newacronym{EVD}{EVD}{eigenvalue decomposition}
\newacronym{LDT}{LDT}{Lagrangian Dual Transform}
\newacronym{QT}{QT}{Quadratic Transform}
\newacronym{CDF}{CDF}{cumulative distribution function}
\newacronym{AP}{AP}{access point}
\newacronym{SINR}{SINR}{signal to interference-plus-noise ratio}
\newacronym{SIC}{SIC}{successive interference cancellation}
\newacronym{CSI}{CSI}{channel state information}
\newacronym{LoS}{LoS}{line-of-sight}
\newacronym{NLoS}{NLoS}{non-LoS}
\newacronym{MIMO}{MIMO}{multiple-input multiple-output}
\newacronym{MISO}{MISO}{multiple-input single-output}
\newacronym{SIMO}{SIMO}{single-input multiple-output}
\newacronym{SISO}{SISO}{single-input single-output}
\newacronym{MU}{MU}{multi-user}
\newacronym{JCAS}{JCAS}{joint communication and sensing}
\newacronym{JCR}{JCR}{joint communications and radar}
\newacronym{ISAC}{ISAC}{integrated sensing and communications}
\newacronym{3D}{3D}{three-dimensional}
\newacronym{2D}{2D}{two-dimensional}
\newacronym{1D}{1D}{one-dimensional}
\newacronym{ROI}{ROI}{region of interest}
\newacronym{mmWave}{mmWave}{millimeter-wave}
\newacronym{MF}{MF}{matched-filter}
\newacronym{SotA}{SotA}{state-of-the-art}
\newacronym{AWGN}{AWGN}{additive white Gaussian noise}
\newacronym{BS}{BS}{base station}
\newacronym{UE}{UE}{user equipment}
\newacronym{wlg}{w.l.g.}{without loss of generality}
\newacronym{CLT}{CLT}{central limit theorem}
\newacronym{PDF}{PDF}{probability density function}
\newacronym{ICI}{ICI}{inter-carrier interference}
\newacronym{BER}{BER}{bit error rate}
\newacronym{DoF}{DoF}{degrees-of-freedom}
\newacronym{VGA}{VGA}{vector Gaussian approximation}
\newacronym{FD}{FD}{full-duplex}
\newacronym{FP}{FP}{fractional programming}
\newacronym{CC}{CC}{communication-centric}
\newacronym{RC}{RC}{raised-cosine}
\newacronym{RRC}{RRC}{root raised-cosine}
\newacronym{6G}{6G}{sixth-generation}
\newacronym{V2X}{V2X}{vehicle-to-everything}
\newacronym{LEO}{LEO}{low-earth orbit}
\newacronym{I/O}{I/O}{input-output}
\newacronym{CE}{CE}{channel estimation}
\newacronym{ICC}{ICC}{integrated communication and computing}
\newacronym{ISCC}{ISCC}{integrated sensing, communications and computing}
\newacronym{PAM}{PAM}{pulse amplitude modulation}
\newacronym{iid}{i.i.d.}{independent and identically distributed}
\newacronym{MEC}{MEC}{mobile edge computing}
\newacronym{REMS}{REMS}{reconfigurable electromagnetic structure}
\newacronym{D-RIS}{D-RIS}{diagonal reconfigurable intelligent surface}
\newacronym{BD-RIS}{BD-RIS}{beyond-diagonal reconfigurable intelligent surface}
\newacronym{RIS}{RIS}{reconfigurable intelligent surface}
\newacronym{RE}{RE}{reflective element}
\newacronym{MRT}{MRT}{maximum ratio transmission}
\newacronym{ZF}{ZF}{zero forcing}
\newacronym{SVD}{SVD}{singular value decomposition}
\newacronym{CGA}{CGA}{conjugate gradient ascent}
\newacronym{QCQP}{QCQP}{quadratic constraint quadratic programming}
\newacronym{MMSE}{MMSE}{minimum mean square error}
\newacronym{RBD-RIS}{RBD-RIS}{reciprocal BD-RIS}
\newacronym{mMIMO}{mMIMO}{massive MIMO}
\newacronym{NRBD-RIS}{NRBD-RIS}{non-reciprocal BD-RIS}
\newacronym{CG}{CG}{conjugate gradient}
\newacronym{B5G}{B5G}{beyond fifth-generation}
\newacronym{pp-ADMM}{pp-ADMM}{partially proximal alternating direction method of multipliers}

\begin{document}
\title{Reciprocal Beyond-Diagonal Reconfigurable Intelligent Surface \!\!\! (BD-RIS):  \,Scattering\,Matrix\,Design\,via\,Manifold\,Optimization \!\!\!}

\title{Fractional Programming and Manifold Optimization\\ Design of Reciprocal BD-RIS Scattering Matrices}

\title{Fractional Programming and Manifold Optimization for\\ Reciprocal BD-RIS Scattering Matrix Design}



\author{\IEEEauthorblockN{Marko Fidanovski,$\!\!^*$ Iv{\'a}n Alexander Morales Sandoval,$\!\!^*$ Kuranage Roche Rayan Ranasinghe,$\!\!^*$ \\ Giuseppe Thadeu Freitas de Abreu,$\!\!^*$ Emil Bj{\"o}rnson$^\dag$ and Bruno Clerckx$^\ddag$}
\IEEEauthorblockA{$^*$\textit{School of Computer Science and Engineering, Constructor University, Bremen, Germany} \\
$^\dag$\textit{School of Electrical Engineering and Computer Science, KTH Royal Institute of Technology, Stockholm, Sweden} \\
$^\ddag$\textit{Department of Electrical and Electronic Engineering, Imperial College London, London, U.K.} \\
Emails: [mfidanovsk, imorales, kranasinghe, gabreu]@constructor.university, emilbjo@kth.se, b.clerckx@imperial.ac.uk}
}

\markboth{To be submitted to the IEEE Wireless Communications Letters, 2025}%
{How to Use the IEEEtran \LaTeX \ Templates}

\maketitle
\begin{abstract}

We investigate the problem of maximizing the sum-rate performance of a \ac{BD-RIS}-aided \ac{MU}-\ac{MISO} system using \ac{FP} techniques.
More specifically, we leverage the \ac{LDT} and \ac{QT} to derive an equivalent objective function which is then solved iteratively via a manifold optimization framework.
It is shown that these techniques reduce the complexity of the optimization problem for the scattering matrix solution, while also providing notable performance gains compared to \ac{SotA} methods under the same system conditions.
Simulation results confirm the effectiveness of the proposed method in improving sum-rate performance.

\end{abstract}


\glsresetall

\section{Introduction}
\IEEEPARstart{I}{ntelligent} reflecting metasurfaces, or \acp{RIS}, are low-power, reconfigurable surfaces that can manipulate the propagation of incident electromagnetic waves to enhance system performance \cite{BjornsonSPM2022,BjornsonNP2024, DiRenzoJSAC2020,LiArx2025}.
As a result of their low-cost and reconfigurable nature, \acp{RIS} -- and more recently \acp{BD-RIS} -- have gained significant attention in the wireless communications research community as a promising technology for enhancing coverage and capacity of \ac{6G} networks \cite{FidanovskiarX2025, YahyaOJCS2024,FangCL2024,WuarX2024,ZhouTWC2024,ZhouCL2025,ZhaoarX2025, LiArx2025_NRBD-RIS,LiTWC2024,LiuArx2025}.

One of the main challenges associated with the deployment of \ac{BD-RIS} is the complexity of the architecture itself; inherently, the more complex the architecture, the higher the potential for performance gains.
An important research direction has been therefore the design of ``optimal" architectures that achieve an effective tradeoff between performance and complexity \cite{LiArx2025}.
Another challenge arises from the need for the scattering matrix to satisfy specific constraints that preserve the passive and reciprocal nature of the structure.
In general, both reciprocal and non-reciprocal \acp{BD-RIS} are physically realizable, where non-reciprocal \acp{BD-RIS} have been shown to provide significant performance gains when compared to their reciprocal counterparts \cite{LiArx2025_NRBD-RIS,LiTWC2024,LiuArx2025}; however, this comes at the cost of higher hardware complexity.

For this reason, most existing studies, including this work, focus on reciprocal \acp{BD-RIS} architectures \cite{FidanovskiarX2025, YahyaOJCS2024,FangCL2024,WuarX2024,ZhouTWC2024,ZhouCL2025,ZhaoarX2025}.
An excellent example is provided in \cite{WuarX2024}, where, among other contributions, the authors proposed an algorithm for the joint design of reciprocal \ac{BD-RIS} scattering and \ac{BS} beamforming matrices to maximize the sum-rate performance in \ac{MU}-\ac{MISO} systems.
To this extent, a general solution applicable to arbitrary architectures was provided, incorporating both the unitary and symmetry constraints with the help of auxiliary variables following a \ac{pp-ADMM} framework.
Additionally, a performance-complexity trade-off analysis was conducted across the fully-, group-, tree-, and single-connected \ac{BD-RIS} architectures.
All in all, the optimal parametrization of \ac{BD-RIS} is a challenging problem which includes, besides the aforementioned issues, also problems such as Pareto tradeoff for single-user, \ac{MU}-\ac{MIMO}, under both lossy and lossless \ac{RIS} architectures, which have been thoroughly studied in current literature.
For more on these issues we refer the reader to \cite{LiArx2025} and references thereby.

This paper serves as a direct improvement of \cite{FidanovskiarX2025}, providing a solution for the fully-, group-, and single-connected architectures\footnote{Extending the framework to additional architectures would require a detailed reformulation of the problem, particularly regarding the scattering matrix constraints, and is therefore left for future research.}.
In that approach, the reciprocal scattering matrix is obtained following a manifold optimization framework\!\,\cite{LiTWC2023} without the use of auxiliary variables to enforce the constraints. 
A key limitation of \cite{FidanovskiarX2025}, however, is that the considered sum-rate function is inherently non-convex, making the optimization prone to convergence to local optima, thus degrading overall performance.
As the main contribution of this article, the solution for the scattering matrix from \cite{FidanovskiarX2025} was enhanced by applying \ac{FP} techniques to the sum-rate function, while maintaining a manifold optimization framework and enforcing symmetry through a penalty term in the objective function.
Finally, using the projection method from \cite{FangCL2024}, the solution is projected onto the feasible set of scattering matrices, namely, unitary and symmetric matrices.
To showcase the improvement in complexity and performance, both analytical and numerical results are provided.

\section{System Model}
\label{sec:sysmodel}


The system model follows the formulation established in \cite{FidanovskiarX2025}, considering a \ac{BD-RIS}-aided downlink \ac{MU}-\ac{MISO} system, as illustrated in Figure~\ref{fig:system model}.
In this setup, a \ac{BS} equipped with $N$ \ac{TX} antennas serves $K$ single \ac{RX} antenna users, denoted as $U_k$, $\forall k \in \{1,2,\dots,K\}$, assisted by a \ac{BD-RIS} composed of $R$ \acp{RE}.
The \ac{BD-RIS} scattering matrix is represented by $\mathbf{\Theta} \in \mathbb{C}^{R\times R}$. 
Furthermore, the channel connecting the \ac{BS} and the \ac{BD-RIS} is referred to as the \ac{BS}-\ac{BD-RIS} channel, and is denoted by $\mathbf{H}_{\mathrm{TX}} \in \mathbb{C}^{R \times N}$.
Similarly, the \ac{BD-RIS}-$U_k$ channel is denoted as $\mathbf{h}_k \in \mathbb{C}^{R \times 1}$.

The transmit signal vector is represented as $\mathbf{x}=\mathbf{V}\mathbf{s}$, such that the information symbols $\mathbf{s} \in \mathbb{C}^{K \times 1}$ must satisfy $\mathbb{E}[\mathbf{s}\mathbf{s}^H]=\mathbf{I}$, and the beamforming matrix $\mathbf{V} \in \mathbb{C}^{N \times K}$ must meet the power constraint $\|\mathbf{V}\|_F^2 \leq P_{\mathrm{max}}$, with $P_{\mathrm{max}}$ being the maximum transmit power at the \ac{BS}. 
To focus on the most challenging and conceptually relevant scenario where the \ac{BD-RIS} plays a critical role, the direct \ac{LoS} link between the \ac{BS} and each user $U_k$ is assumed to be blocked.
As a result, $U_k$ receives the signal $r_k \in \mathbb{C}$, which can be expressed as
\vspace{-1ex}
\begin{equation}
\label{eq:rsig}
r_k = \mathbf{h}_k^T\mathbf{\Theta}\mathbf{H}_{\mathrm{TX}}\mathbf{x}+n_k,
\end{equation}
where $n_k \sim \mathcal{CN}(0, N_0)$ denotes \ac{AWGN} with power $N_0$. 

For convenience, the notation used in \cite{FidanovskiarX2025,YahyaOJCS2024} is respected, wherein the compact vector representation is introduced as
\begin{subequations}
\begin{gather}
\mathbf{r} = [r_1, r_2, \dots, r_K]^T, \\
\mathbf{n} = [n_1, n_2, \dots, n_K]^T, \\
\mathbf{V} = [\mathbf{v}_1, \mathbf{v}_2, \dots, \mathbf{v}_K], \\
\mathbf{H}_{\mathrm{RX}} = [\mathbf{h}_1, \mathbf{h}_2, \dots, \mathbf{h}_K]^T,  \\
\mathbf{H}_{\mathrm{TX}}= \left[ \mathbf{w}_1, \mathbf{w}_2, \dots, \mathbf{w}_N \right], \\ 
\mathbf{r} = \mathbf{H}_{\mathrm{RX}} \boldsymbol{\Theta} \mathbf{H}_{\mathrm{TX}} \mathbf{x} + \mathbf{n}, \label{eq:rvec}
\end{gather}
\end{subequations}
with $\mathbf{r}\in\mathbb{C}^{K\times 1}$, $\mathbf{H}_{\mathrm{RX}} \in \mathbb{C}^{K \times R}$, and $\mathbf{w}_n \in \mathbb{C}^{R\times 1}$, s.t. $n\in \{1,2,\dots,N\}$, denoting the received signal vector, \ac{BD-RIS}-$U_k$ channel matrix, and \ac{BS}-\ac{BD-RIS} channel vector, respectively.

\vspace{-1ex}
\subsection{Scattering Matrix Definition}

For the sake of direct comparison, the same three \ac{BD-RIS} architectures as in \cite{FidanovskiarX2025} are considered.
These architectures differ based on the configuration of the reconfigurable impedance network and must satisfy the following constraints to ensure the reciprocal and passive properties of the structure, namely:
\begin{enumerate}
\item Single-connected \ac{BD-RIS}: Simplest architecture considered which is equivalent to the conventional D-RIS.
The aforementioned constraints are described as 
\begin{eqnarray}
&\mathcal{S}_{\text{SC}_1} = \left\{ \mathbf{\Theta} : [\mathbf{\Theta}]_{i,j} = 0,\ \forall i \ne j \right\},& \label{eq:SC_sym}\\
&\mathcal{S}_{\text{SC}_2} = \left\{ \mathbf{\Theta} : \left|[\mathbf{\Theta}]_{i,j}\right| = 1,\ \forall i = j \right\},&
\end{eqnarray}
where $i,j \in \{1, 2, \dots, R\}$.  

\item Fully-connected \ac{BD-RIS}: The corresponding scattering matrix of this architecture is full and -- similarly to the other architectures-- must satisfy both the symmetry and the unitary constraint, which are given as

\begin{figure}[H]
\centering
\includegraphics[width=1\linewidth]{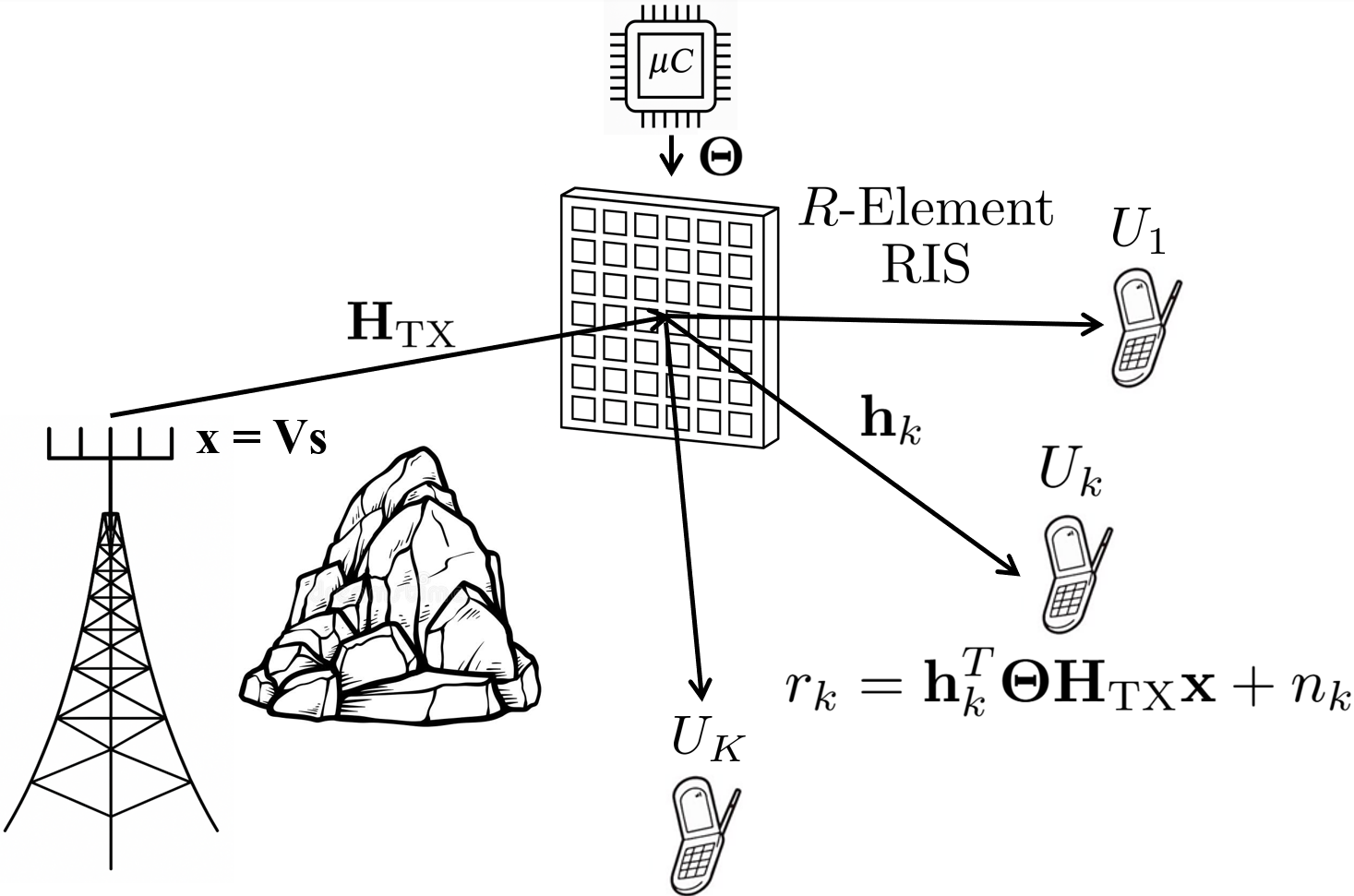}
\caption{Illustration of the system model, where a \ac{BS} with $N$ \ac{TX} antennas serves $K$ single-antenna users through an $R$-Element \ac{RBD-RIS}, without a \ac{LoS} link between the BS and the users.}
\label{fig:system model}
\vspace{-2ex}
\end{figure}
\vspace{-4ex}
\begin{eqnarray}
&\mathcal{S}_{\text{FC}_1} = 
\left\{ 
  \mathbf{\Theta} : 
  \mathbf{\Theta} = \mathbf{\Theta}^T 
\right\},&\\
&\mathcal{S}_{\text{FC}_2} = 
\left\{ 
\mathbf{\Theta} : 
\mathbf{\Theta} \mathbf{\Theta}^H = \mathbf{I} 
\right\}.&
\end{eqnarray}

\item Group-connected \ac{BD-RIS}: This intermediate architecture represents a general case.
The \acp{RE} are divided into $G$ disjoint groups, each consisting of $R_G = \frac{R}{G}$ elements.
This configuration reduces to the fully-connected case when $G=1$ and to the single-connected case when $G=R$.
Each group of the scattering matrix is denoted by $\mathbf{\Theta}_g \in \mathbb{C}^{R_G \times R_G}$, where $g \in \{1, 2, \ldots, G\}$.
Accordingly, the constraints on the group-connected BD-RIS scattering matrix can be described as
\begin{gather}
\hspace{-5ex} \mathcal{S}_{\text{GC}_1} \!\!=\!
\left\{ 
  \mathbf{\Theta} \!:\! 
  \mathbf{\Theta} \!=\! \mathrm{blkdiag}(\mathbf{\Theta}_1, \ldots, \mathbf{\Theta}_G), \! 
  \mathbf{\Theta}_g \!=\! \mathbf{\Theta}_g^T, \forall g
\right\}\!,\!\!\!
\end{gather}
\begin{equation}
\mathcal{S}_{\text{GC}_2} = 
\left\{ 
  \mathbf{\Theta} : 
  \mathbf{\Theta}_g \mathbf{\Theta}_g^H = \mathbf{I}_{R_G},\; \forall g 
\right\}. \label{eq:GC_uni}
\end{equation}
\end{enumerate}

For further details on the architectures, we refer the reader to \cite{LiArx2025, FidanovskiarX2025, YahyaOJCS2024}.

\vspace{-2ex}
\subsection{Problem Formulation}

Given $\mathbf{\Theta}$ and $\mathbf{V}$, and assuming perfect channel state information, an achievable rate at $U_k$ is $\log_2(1+\gamma_k)$, where the \ac{SINR} is expressed as
\vspace{-1ex}
\begin{equation}
\gamma_k = 
\frac{ \left| \mathbf{h}_k^T \mathbf{\Theta} \mathbf{H}_{\mathrm{TX}} \mathbf{v}_k \right|^2 }
{ \sum\limits_{i \ne k} \left| \mathbf{h}_k^T \mathbf{\Theta} \mathbf{H}_{\mathrm{TX}} \mathbf{v}_i \right|^2 + N_0 }
.
\vspace{-1ex}
\end{equation}

The corresponding sum-rate maximization problem is then
\vspace{-0.5ex}
\begin{subequations} \label{prob:P1}
\begin{align}
(\text{P1}): \quad \underset{\mathbf{V},\mathbf{\Theta}}{\mathrm{maximize}} \quad & \sum_k \log_2(1 + \gamma_k) \label{eq:P1_obj} \\
\text{subject to} \quad & \|\mathbf{V}\|_F^2 \leq P_{\mathrm{max}}, \label{eq:P1_power} \\
& \mathbf{\Theta} \in \mathcal{S}_{a_1}, \label{eq:P1_Sa1} \\
& \mathbf{\Theta} \in \mathcal{S}_{a_2}, \label{eq:P1_Sa2}
\end{align}
\end{subequations}
where $a \in \{ \text{SC}, \text{FC}, \text{GC} \}$, as described in \eqref{eq:SC_sym}-\eqref{eq:GC_uni}.

Since the aforementioned limitation is only on the scattering matrix design stage, this paper focuses on solving (P1) with respect to $\mathbf{\Theta}$ following \cite{FidanovskiarX2025}, while the design of the beamforming matrix $\mathbf{V}$ will be presented in a follow-up work.

\section{Stage 1: Scattering Matrix Design}
\label{sec:smdesign}

The optimization problem (P1) is solved following an iterative process.
To design the scattering matrix, the impact of the \ac{BS} beamforming is fixed by initializing it with a \ac{MMSE}-based beamformer. 
Accordingly, \eqref{eq:rvec} is rewritten as
\begin{equation}
\mathbf{r}=\mathbf{H}_{\mathrm{RX}}\mathbf{\Theta}\mathbf{H}_{\mathrm{TX}}\mathbf{V}\mathbf{s}+\mathbf{n}
\triangleq \mathbf{H}_{\mathrm{RX}}\mathbf{\Omega}\mathbf{V}\mathbf{s}+\mathbf{n}
\triangleq \mathbf{E}\mathbf{V}\mathbf{s}+\mathbf{n},
\end{equation}
where $\mathbf{\Omega} = \mathbf{\Theta}\mathbf{H}_{\mathrm{TX}} = [\boldsymbol{\omega}_1,\dots, \boldsymbol{\omega}_N] \in \mathbb{C}^{R\times N}$ and $\mathbf{E} = \mathbf{H}_{\mathrm{RX}}\mathbf{\Omega}= [\mathbf{e}_1,\dots,\mathbf{e}_K]^T \in \mathbb{C}^{K\times N}$ denotes the equivalent channel.

Correspondingly, the \ac{SINR} at $U_k$ during the design of the scattering matrix can be rewritten as a function of the equivalent channel, given by
\begin{equation}
\label{eq:SINR}
\hspace{-1ex}
\gamma_k = 
\frac{ \left| \mathbf{h}_k^T \boldsymbol{\Omega} \mathbf{v}_k\right|^2}
{\sum\limits_{i \ne k} \left| \mathbf{h}_k^T \boldsymbol{\Omega} \mathbf{v}_i \right|^2 + N_0}
=
\frac{\left| \mathbf{e}_{k} \mathbf{v}_k \right|^2}
{\sum\limits_{i \ne k} \left| \mathbf{e}_{k} \mathbf{v}_i \right|^2 + N_0},
\end{equation}
where $\mathbf{v}_l \in \mathbb{C}^{N \times 1}$ denotes the beamforming vector, with $l \in\{i,k\}$.

Making use of \eqref{eq:SINR}, the optimization problem (P1) is reformulated as
\begin{subequations} \label{prob:P2}
\begin{align}
  (\text{P2}): \quad \underset{\mathbf{\Theta}}{\mathrm{maximize}} \quad & \sum_k \log_2(1 + \gamma_k)  \\
  \text{subject to} \quad & \mathbf{\Theta} \in \mathcal{S}_{a_1},  \\
  & \mathbf{\Theta} \in \mathcal{S}_{a_2}.
\end{align}
\end{subequations}

\subsection{Solution for the Group-Connected architecture}

We focus on solving (P2) for the group-connected \ac{BD-RIS} architecture since the single-connected and fully-connected architectures are special cases of it. Hence, we rewrite (P2) as
\begin{subequations}  
\begin{align}
 (\text{P2a}): \quad \underset{\mathbf{\Theta}}{\mathrm{maximize}} \quad & \sum_k \log_2(1 + \gamma_k)  \\
\text{subject to} \quad & \mathbf{\Theta}_g = \mathbf{\Theta}_g^T,  \\
 & \mathbf{\Theta}_g\mathbf{\Theta}_g^H = \mathbf{I}_{R_G},
\end{align}
\end{subequations} 
where $g=\{1,2,\dots,G\}$. 

For simplicity, the sum-rate over all users is denoted as
\begin{equation}
\eta = \sum_k \eta_k, \quad \text{with} \quad \eta_k = \log_2(1 + \gamma_k).
\end{equation}

Following the formulation in \cite{FidanovskiarX2025}, the problem (P2a) is solved using manifold optimization techniques.
To this end, the linear subspace, and thus the manifold, induced by the symmetry constraint is implicitly enforced through a penalty term in the objective function.
Moreover, the unitary constraint is directly addressed by treating $\mathbf{\Theta}_g$ as a point on the Stiefel manifold.
Accordingly, the optimization problem then becomes
\begin{subequations}  
\begin{align}
\hspace{-1.35ex}
 (\text{P3}): \quad \underset{\mathbf{\Theta}}{\mathrm{maximize}} \quad & \!\!\sum_k \log_2(1 + \gamma_k)
  - \nu \bigl\| \boldsymbol{\Theta} - \boldsymbol{\Theta}^T \bigr\|_F^2
\label{eq:P3_obj}\\
\text{subject to} \quad & \!\!\mathbf{\Theta}_g\mathbf{\Theta}_g^H = \mathbf{I}_{R_G},\label{eq:P3_const}
\end{align}
\end{subequations}
where $\nu \in \mathbb{R}$ denotes a nonnegative weight\footnote{The optimization of this parameter is left for detailed investigation in the journal extension.} and the term $\bigl\| \boldsymbol{\Theta} - \boldsymbol{\Theta}^T \bigr\|_F^{2}
$ is used to obtain a quantitative measure of how far the matrix is from being symmetric.

Furthermore, (P3) can be simplified by applying scalar \ac{FP} techniques \cite{ShenTSP2018_I,ShenTSP2018_II,shen2018phd}, which transform the sum-rate objective into a convex form.
Particularly, we apply the \ac{LDT}, which yields the equivalent problem
\begin{equation}
  \bar{\eta}_k = \log_2(1+\tau_k) - \frac{\tau_k}{\ln(2)} + \frac{1+\tau_k}{\ln(2)}\cdot\frac{\left| \mathbf{e}_{k} \mathbf{v}_k \right|^2}
{\sum\limits_{i} \left| \mathbf{e}_{k} \mathbf{v}_i \right|^2 + N_0},
\end{equation}
where the auxiliary variable $\tau_k \in \mathbb{C}$ denotes the Lagrange multiplier, s.t. $\tau_k = \gamma_k$. 

Furthermore, the \ac{QT} is applied to the fractional term in $\bar{\eta}_k$, yielding the equivalent form 
\begin{align}
  \label{eq:eta_hat}
  \hat{\eta}_k = &  \log_2(1+\tau_k) - \frac{\tau_k}{\ln(2)} \\
  & + \frac{1+\tau_k}{\ln(2)} \Big[ 2\Re\{y_k^{\star}\mathbf{e}_k\mathbf{v}_k\}-\left|y_k\right|^2\big(\sum_{i}\left|\mathbf{e}_k\mathbf{v}_i\right|^2+N_0\big)\Big], \nonumber
\end{align}
where $y_k \in \mathbb{C}$ is an auxiliary variable corresponding to the \ac{QT}, given as
\begin{equation}
  y_k = \frac{\mathbf{e}_k\mathbf{v}_k}{\sum\limits_{i}\left|\mathbf{e}_k\mathbf{v}_i\right|^2+N_0}.
\end{equation}

In that way, the optimization problem (P3) can be reformulated as
\begin{subequations}  
\begin{align}
\hspace{-1.35ex}
 (\text{P3a}): \quad \underset{\mathbf{\Theta}}{\mathrm{maximize}} \quad & \!\!\sum_k \hat{\eta}_k
  - \nu \bigl\| \boldsymbol{\Theta} - \boldsymbol{\Theta}^T \bigr\|_F^2
\label{eq:P3a_obj}\\
\text{subject to} \quad & \!\!\mathbf{\Theta}_g\mathbf{\Theta}_g^H = \mathbf{I}_{R_G},\label{eq:P3a_const}
\end{align}
\end{subequations}
where for simplicity, the modified sum-rate objective function in \eqref{eq:P3a_obj} including the penalty term is denoted as $\breve{\eta}_k$.

The solution to (P3a) can be obtained numerically using the Manopt toolbox \cite{BoumalJMLR2014}; however, this approach incurs a high computational cost.
To address this, we modify the \ac{CGA} algorithm proposed in \cite{FidanovskiarX2025} to include the computation of the auxiliary variables which stem from the \ac{FP} transformations.
This algorithm utilizes a closed-form expression for the gradient of the objective function, as observed in \eqref{eq:P3a_obj}, which enables a more efficient implementation of the scattering matrix design.

To derive the gradient for the general case, it is necessary to express the system model in a group-wise manner, namely
\begin{equation}
\mathbf{\Theta}_g = \mathbf{\Theta}_{[R_G(g-1)+1\,:\,gR_G,\;\; R_G(g-1)+1\,:\,gR_G]},
\end{equation}
where $g \in \{1,2,\dots,G\}$.

Accordingly, the group-wise equivalent channel is defined as $\mathbf{e}_k^{(g)} = \mathbf{h}_k^{(g)T} \mathbf{\Theta}_g \mathbf{W}^{(g)}$, where $\mathbf{e}_k^{(g)} \in \mathbb{C}^{1\times N}$, the vector $\mathbf{h}_k^{(g)T} = \mathbf{H}_{RX\left[k, R_G(g-1)+1:gR_G\right]}$ and the matrix $\mathbf{W}^{(g)} = \mathbf{H}_{TX\left[R_G(g-1)+1:gR_G, 1:N\right]}$ represents the channel components associated with the $g$-th group.
\clearpage


\quad\\[-5ex]

\begin{algorithm}[H]
\caption{Proposed CGA for BD-RIS Optimization}
\label{alg:cga}
\begin{algorithmic}[1]
\REQUIRE $\mathbf{H}_{\mathrm{TX}}, \mathbf{H}_{\mathrm{RX}}, P_{\mathrm{max}}, N, N_0, G, \nu, I, \epsilon$
\ENSURE Optimized RIS matrix $\mathbf{\Theta}^{\mathrm{opt}}$
\INITIALIZE $\mathbf{\Theta}^{(0)}$ as a random block-diagonal unitary symmetric matrix
\STATE Compute initial FP auxiliary variables $\tau_k$ and $y_k$, $\forall k$
\STATE Compute initial objective $\breve{\eta}(\mathbf{\Theta}^{(0)})$ and Riemannian gradient $\mathbf{r}^{(0)} = \mathrm{T}_{\mathbf{\Theta}}(\nabla_{\mathbf{\Theta}}\breve{\eta}(\mathbf{\Theta}^{(0)}), \mathbf{\Theta}^{(0)})$
\vspace{0.125em}
\STATE Set initial search direction $\boldsymbol{\Xi}^{(0)} = -\mathbf{r}^{(0)}$
\FOR{$i = 0$ to $I$}
\IF{$\langle \mathbf{r}^{(i)}, \boldsymbol{\Xi}^{(i)} \rangle \leq 0$}
\vspace{0.125em}
    \STATE Set $\boldsymbol{\Xi}^{(i)} = \mathbf{r}^{(i)}$
\ENDIF
\STATE Compute $\alpha^{(i)}$ via Armijo line search and update $\mathbf{\Theta}^{(i+1)}$
\vspace{-2.5ex}
\STATE Compute new FP auxiliary variables $\tau_k$ and $y_k$, $\forall k$
\STATE Compute new objective $\breve{\eta}(\mathbf{\Theta}^{(i+1)})$, {using \eqref{eq:P3a_obj}}
\STATE Compute new Riemannian gradient $\mathbf{r}^{(i+1)}$, using \eqref{eq:tang_proj}
\vspace{0.125em}
\STATE Compute $\beta^{(i)} = \max\left(0, \frac{\langle \mathbf{r}^{(i+1)}, \mathbf{r}^{(i+1)} - \mathbf{r}^{(i)} \rangle}{\langle \mathbf{r}^{(i)}, \boldsymbol{\Xi}^{(i)} \rangle}\right)$
\vspace{0.125em}
\STATE Update direction: $\boldsymbol{\Xi}^{(i+1)} = -\mathbf{r}^{(i+1)} + \beta^{(i)} \boldsymbol{\Xi}^{(i)}$
\IF{$|\eta(\mathbf{\Theta}^{(i+1)}) - \eta(\mathbf{\Theta}^{(i)})| < \epsilon$}
    \STATE \textbf{break}
\ENDIF
\ENDFOR
\FOR{$g = 1$ to $G$}
\STATE Extract block $\mathbf{\Theta}_g$ from $\mathbf{\Theta}^{(i+1)}$
\STATE Symmetrize: $\mathbf{\Theta}_{\mathrm{sym}} = \frac{1}{2}(\mathbf{\Theta}_g + \mathbf{\Theta}_g^{T})$
\STATE Perform SVD: $\mathbf{\Theta}_{\mathrm{sym}} = \mathbf{U}\mathbf{\Sigma}\mathbf{V}^H$
\STATE Set $\mathbf{\Theta}^{\mathrm{opt}}_{g} = \mathbf{U}\mathbf{V}^H$
\ENDFOR
\STATE Assemble full $\mathbf{\Theta}^{\mathrm{opt}}$ as block diagonal of all $\mathbf{\Theta}_{g}^{\mathrm{opt}}$
\RETURN $\mathbf{\Theta}_{\mathrm{opt}}$
\end{algorithmic}
\end{algorithm}

\vspace{-2ex}
As such, the gradient of the objective function \eqref{eq:P3a_obj} with respect to each group $\mathbf{\Theta}_g$ is computed as
\begin{equation}
\label{eq:grad}
\nabla_{\mathbf{\Theta}_g}\!\breve{\eta}  =  \nabla_{\mathbf{\Theta}_g} \! \Big( \!  \sum_k \hat{\eta}_k - \nu \left\| \mathbf{\Theta}_g - \mathbf{\Theta}_g^T \right\|_F^2 \! \Big),
\vspace{-1ex}
\end{equation}
where the gradient of $\left\| \mathbf{\Theta}_g - \mathbf{\Theta}_g^T \right\|_F^2$ with respect to $\mathbf{\Theta}_g$ is derived as
\begin{equation}
\nabla_{\mathbf{\Theta}_g}\left\| \mathbf{\Theta}_g - \mathbf{\Theta}_g^T \right\|_F^2 = 4\left(\mathbf{\Theta}_g-\mathbf{\Theta}_g^T\right).
\end{equation}

Omitting the constant terms of the equivalent sum-rate objective after the \ac{FP} transformations allows for the gradient to be expressed as
\begin{equation}
\label{eq:gradD}
\nabla_{\mathbf{\Theta}_g} \hat{\eta}_k \!=\!  \nabla_{\mathbf{\Theta}_g} \!\Big( 2\Re\{y_k^{\star}\mathbf{e}_k\mathbf{v}_k\}\!-\!\left|y_k\right|^2\!\big(\sum_{i}\left|\mathbf{e}_k\mathbf{v}_i\right|^2\!+\!N_0\big) \Big),
\end{equation}
where the gradient with respect to both terms is given as 
\begin{equation}
  \nabla_{\mathbf{\Theta}_g} 2\Re\{y_k^{\star}\mathbf{e}_k\mathbf{v}_k\} = 2\left(y_k^{\star}\mathbf{h}_k^{(g)}\big(\mathbf{W}^{(g)}\mathbf{v}_k \big)^T\right)^{\star},
  \end{equation}
and
\begin{align}
  \nabla_{\mathbf{\Theta}_g} \left|y_k\right|^2\big(&\sum_{i}\left|\mathbf{e}_k\mathbf{v}_i\right|^2+N_0\big) = \\[-1ex]
   &2 \left|y_k\right|^2\sum_{i}\left(\big(\mathbf{e}_k\mathbf{v}_i\big)^{\star}\mathbf{h}_k^{(g)}\big(\mathbf{W}^{(g)}\mathbf{v}_i\big)^{T}\right)^{\star}. \nonumber
\end{align}

\begin{figure}[H]
    \centering
    \includegraphics[width=1\linewidth]{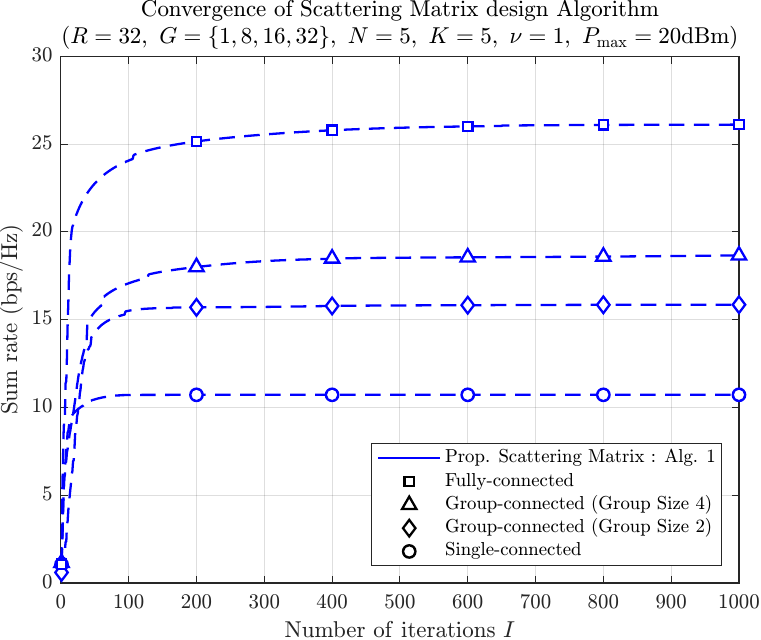}
    \caption{Convergence of Algorithm 1 vs. number of iterations $I$ for considered \ac{BD-RIS} architectures.}
    \label{fig:convergence}
\end{figure}

We refer the reader to \cite{FidanovskiarX2025} for further details on the derivation of the gradient.
Accordingly, the gradient of the objective function \eqref{eq:grad} is reformulated as shown in \eqref{eq:fullgradbf}.
For convenience, \eqref{eq:fullgrad} presents the gradient of the objective function assuming a simple power allocation matrix used for beamforming\footnotemark, $i.e.$, $\mathbf{V} = \mathrm{diag}(\sqrt{v}_1, \dots, \sqrt{v}_K)$, where $v_k \in \mathbb{R}, \forall k$.
Both gradients are provided in closed-form at the top of the next page.

\footnotetext{This assumption can only be made for fully-loaded systems ($K$ = $N$).}

\begin{figure*}[t]
\begin{align}
\nabla_{\mathbf{\Theta}_g}\breve{\eta} =  \sum_k \frac{1+\tau_k}{\ln(2)}\Big[2\left(y_k^{\star}\mathbf{h}_k^{(g)}\big(\mathbf{W}^{(g)}\mathbf{v}_k \big)^T\right)^{\star} - 2 \left|y_k\right|^2\sum_{i}\left(\big(\mathbf{e}_k\mathbf{v}_i\big)^{\star}\mathbf{h}_k^{(g)}\big(\mathbf{W}^{(g)}\mathbf{v}_i\big)^{T}\right)^{\star}\Big] \!- \!
    4\left(\mathbf{\Theta}_g-\mathbf{\Theta}_g^T\right).
\label{eq:fullgradbf}
\end{align}
\begin{center}
\rule{0.5\textwidth}{0.1pt}
\begin{align}
\nabla_{\mathbf{\Theta}_g}\breve{\eta} = \sum_k \frac{1+\tau_k}{\ln(2)}\Big[2v_k\big(y_k^{\star}\mathbf{h}_k^{(g)}\mathbf{w}_k^{(g)T}\big)^{\star} -2\left|y_k\right|^2\sum_{i}v_i^2\big(e_{k,i}^{\star}\mathbf{h}_k^{(g)}\mathbf{w}_i^{(g)T}\big)^{\star}\Big] - 4\nu(\mathbf{\Theta}_g \!- \mathbf{\Theta}_g^T).
\label{eq:fullgrad}
\end{align}
\rule{1\textwidth}{0.1pt}
\end{center}
\vspace{-2ex}
\end{figure*}


The main operations required to implement the \ac{CGA} algorithm are the retraction $\mathrm{R}\big(\cdot,\cdot\big)$ and tangent projection $\mathrm{T}\big(\cdot,\cdot\big)$ functions, which are defined as 
\begin{equation}
\label{eq:retract}
\mathrm{R}_\mathbf{\Theta}(\mathbf{\Theta}, \boldsymbol{\Xi}) = \mathcal{Q}(\mathbf{\Theta} + \alpha \boldsymbol{\Xi}),
\end{equation}
and
\vspace{-1ex}
\begin{equation}
\label{eq:tang_proj}
\mathrm{T}_{\mathbf{\Theta}}( \nabla_{\mathbf{\Theta}} \eta , \mathbf{\Theta}) = \nabla_{\mathbf{\Theta}} \eta - \mathbf{\Theta} \cdot \frac{\mathbf{\Theta}^H\nabla_{\mathbf{\Theta}} \eta + (\nabla_{\mathbf{\Theta}} \eta )^H \mathbf{\Theta}}{2},
\vspace{-0.5ex}
\end{equation}
where $\alpha$ denotes the step size, $\mathbf{\Theta}$ the current point on the manifold, and $\boldsymbol{\Xi}$ the ascent direction in the tangent space.

Furthermore, $\mathcal{Q}\big(\cdot\big)$ denotes an operator which returns the Q-factor from the QR decomposition of a given matrix.
Without loss of generality, both the retraction and tangent projection functions are defined for each block $\mathbf{\Theta}_g$.

The full procedure for solving (P3a) using the aforementioned steps is summarized in Algorithm \ref{alg:cga}. 
For a detailed description, the reader is referred to \cite{FidanovskiarX2025}.

\textit{Computational Complexity:} The computational complexity of the proposed Algorithm \ref{alg:cga} is very similar to that of the algorithm in \cite{FidanovskiarX2025} with the main differences being that the gradient in \eqref{eq:fullgradbf} is much simpler, yielding a complexity of $\mathcal{O}(K^2 G (R/G)^2)$, and the computation of the \ac{FP} auxiliary variables which further amount to $\mathcal{O}(K^2N)$.
For clarity, the convergence behavior of Algorithm \ref{alg:cga} for each considered \ac{BD-RIS} architecture (single-, group- with group sizes of 2 and 4, and fully connected) is illustrated in Figure \ref{fig:convergence}.

Furthermore, unique markers are used to represent each architecture, as described in the legend of Figure \ref{fig:convergence}, and this notation is applied consistently throughout the article.

The simulation setup adopts the same parameter configuration as in \cite{FidanovskiarX2025}.
In particular, we consider the following parameters: convergence tolerance is set to $\epsilon = 10^{-8}$, maximum maximum number of \ac{CGA} iterations $I = 8000$, maximum Armijo line search steps $L = 200$, sufficient increase coefficient for the stepsize of $2\times 10^{-11}$, initial stepsize $c_{\text{init}} = 1$, contraction factor for the stepsize $c_{\text{dec}} = 0.75$, and the symmetry enforcing penalty constant $\nu = 1$. 
A consistent trend is observed where architectures with higher connectivity require more iterations to reach convergence.
Notably, Algorithm \ref{alg:cga} exhibits much faster convergence compared to the results shown in \cite[Fig.~2]{FidanovskiarX2025}.

This improvement is consistent with the comparable per-iteration computational complexity derived earlier in the paper, and the lower number of iterations required to reach convergence across all \ac{BD-RIS} architectures.
Specifically, \cite{FidanovskiarX2025} reports convergence after $50$ iterations for the single-connected case, $\sim700$ and $\sim1000$ for the group-connected architecture with a group size of 2 and 4, respectively, and $\sim 2500$ iterations for the fully-connected structure.
In contrast, the proposed method converges in approximately $50, 100, 400,$ and $700$ iterations, respectively, considering the same \ac{BD-RIS} configurations.


%
%

\vspace{-1ex}
\section{Simulation Results}
\label{sec:simresults}

This section aims to validate the effectiveness of the proposed algorithm in enhancing communications performance compared to existing methods for scattering matrix design \cite{YahyaOJCS2024, WuarX2024}.
The evaluation is carried out through computer simulations, such that the relevant parameters to be directly observed from the figures presented.
%
Finally, the same channel and pathloss models and parameters as in \cite{YahyaOJCS2024} are adopted.





\begin{figure}[H]
  \centering
  \includegraphics[width=\linewidth]{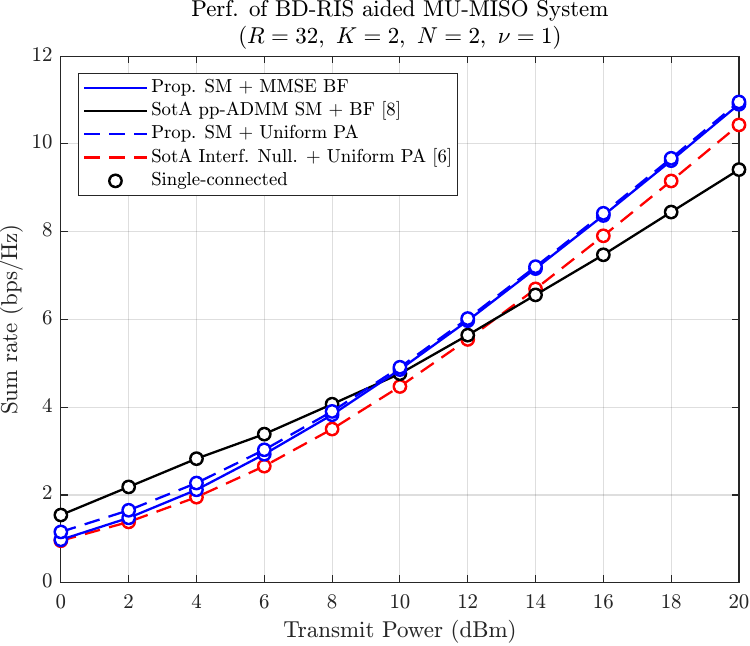}
  \caption{Comparison of sum-rate performance of the proposed vs. SotA \cite{YahyaOJCS2024, WuarX2024} scattering matrix ''SM" design for the single-connected ''SC" architecture.}
  \label{fig:fig_1}
\vspace{4ex}
  \includegraphics[width=\linewidth]{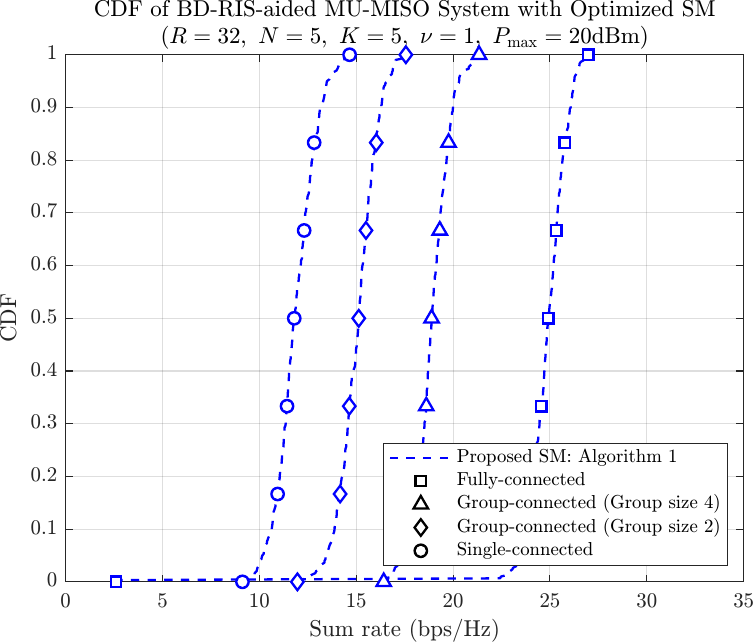}
  \caption{CDF of sum-rate performance of the proposed scattering matrix ``SM" design with uniform power allocation, considering the fully-connected ``FC", group-connected, with group sizes of 2 and 4, ``GC(2)", and ``GC(4)", and the single-connected ``SC" architecture.}
  \label{fig:fig_2}
  \vspace{-2ex}
\end{figure}

Figure \ref{fig:fig_1} presents the sum-rate performance of the proposed scattering matrix design combined with \ac{SotA} beamforming ``BF'' schemes, $i.e.$, uniform power allocation ``PA'', and MMSE BF, compared to the \ac{SotA} designs in \cite{YahyaOJCS2024, WuarX2024} for the single-connected architecture, where \cite{WuarX2024} is considered the best sum rate performing joint ``SM'' and ``BF'' design.  
%
We observe that at sufficiently moderate-to-high SNR, the proposed method, employing simple MMSE beamforming and power allocation, outperforms the system based on a jointly optimized SM and BF design \cite{WuarX2024}.

Complementary, Figure \ref{fig:fig_2} illustrates the \ac{CDF} of the sum-rate for all considered architectures to provide a comprehensive performance comparison.
%
%
%

\begin{figure}[H]
    \centering
    \includegraphics[width=\linewidth]{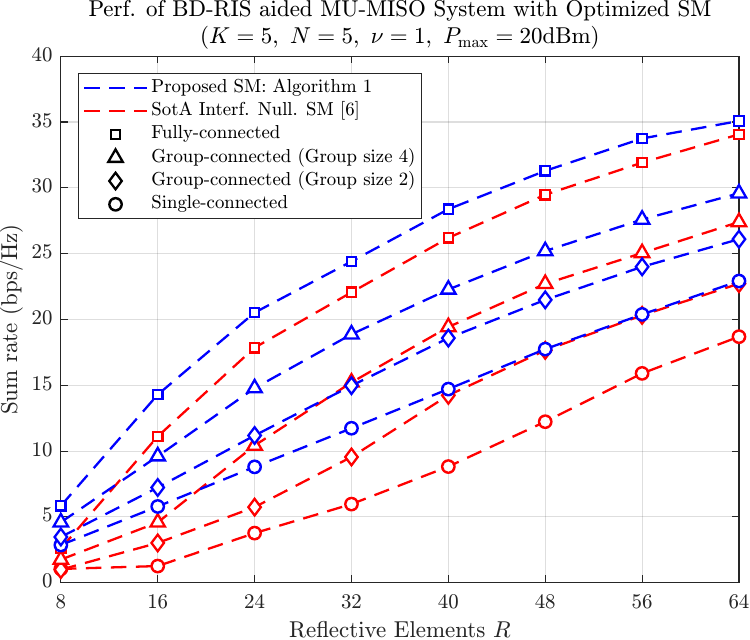}
    \caption{Comparison of sum-rate performance of the proposed vs. \ac{SotA} \cite{YahyaOJCS2024} scattering matrix ``SM" design as a function of the number of \acp{RE} $R$, considering the fully-connected ``FC", group-connected, with group sizes of 2 and 4, ``GC(2)" and ``GC(4)", and the single-connected ``SC" architecture.}
    \label{fig:sRvsRE}
\end{figure}

Lastly, Figure \ref{fig:sRvsRE} demonstrates the sum-rate performance of the proposed method compared with the \ac{SotA} scattering matrix design from \cite{YahyaOJCS2024}, as a function of the number of \acp{RE} $R$.
The evaluation uses the same channel realizations and beamforming scheme, namely uniform power allocation, across all considered \ac{BD-RIS} architectures, including single-, group-, with group size of 2 and 4, and fully-connected configurations.
The proposed method consistently achieves higher sum-rate performance as both the architecture complexity and the number of \acp{RE} increase.
This result further highlights the scalability and robustness of the proposed approach.

\section{Conclusion}
\label{sec:conc}

A novel sum-rate maximization scheme for \ac{RBD-RIS}-aided \ac{MU}-\ac{MISO} systems is presented.
The focus of this article is on the design of the reciprocal scattering matrix, where \ac{FP} techniques are employed to reformulate the objective function into its equivalent convex form.
The resulting problem is then solved using a modified Riemannian gradient descent algorithm, supported by the closed-form expression for the gradient.
The results confirm that, although manifold optimization is well suited for addressing non-convex problems, transforming the formulation into its equivalent convex form, which might not always be possible, yields more reliable and higher-quality solutions.
Overall, the proposed approach achieves notable performance gains while reducing computational complexity, highlighting its potential as a design strategy for future \ac{RIS}-aided wireless systems.

\

\bibliographystyle{IEEEtran}
\bibliography{ref}

\end{document}